\documentclass[conference]{IEEEtran}
\IEEEoverridecommandlockouts
\usepackage{cite}
\usepackage{amsmath,amssymb,amsfonts}
\usepackage{array}
\usepackage{algorithmic, algorithm}
\usepackage{siunitx}
\usepackage{graphicx}
\usepackage{textcomp}
\usepackage{xcolor}
\usepackage{caption}
\usepackage{subcaption}
\def\BibTeX{{\rm B\kern-.05em{\sc i\kern-.025em b}\kern-.08em
    T\kern-.1667em\lower.7ex\hbox{E}\kern-.125emX}}
\begin{document}

\title{A General Framework for Error-controlled Unstructured Scientific Data Compression\\
\thanks{This manuscript has been authored in part by UT-Battelle, LLC, under
contract DE-AC05-00OR22725 with the US Department of Energy (DOE).
The publisher, by accepting the article for publication, acknowledges that
the U.S. Government retains a non-exclusive, paid up, irrevocable, world-
wide license to publish or reproduce the published form of the manuscript, or allow others to do so, for U.S. Government purposes. The DOE will provide
public access to these results in accordance with the DOE Public Access Plan
(http://energy.gov/downloads/doe-public-access-plan).}
}

\author{\IEEEauthorblockN{
Qian Gong\IEEEauthorrefmark{1},
Zhe Wang\IEEEauthorrefmark{1}, 
Viktor Reshniak\IEEEauthorrefmark{1}, 
Xin Liang\IEEEauthorrefmark{2}, 
Jieyang Chen\IEEEauthorrefmark{3}, 
Qing Liu\IEEEauthorrefmark{4}, 
Tushar M. Athawale\IEEEauthorrefmark{1}, 
Yi Ju\IEEEauthorrefmark{6},\\
Anand Rangarajan\IEEEauthorrefmark{5}, 
Sanjay Ranka\IEEEauthorrefmark{5},
Norbert Podhorszki\IEEEauthorrefmark{1}, 
Rick Archibald\IEEEauthorrefmark{1}, 
Scott Klasky\IEEEauthorrefmark{1}
}
\IEEEauthorblockA{{\textit{Oak Ridge National Laboratory}, Oak Ridge, TN}\IEEEauthorrefmark{1} \\
{\textit{University of Kentucky}, Lexington, KY}\IEEEauthorrefmark{2} \\
{\textit{University of Alabama}, Birmingham, AL}\IEEEauthorrefmark{3} \\
{\textit{New Jersey Institute of Technology}, Newark, CA}\IEEEauthorrefmark{4} \\
{\textit{University of Florida}, Gainesville, FL}\IEEEauthorrefmark{5}\\
{\textit{Max Planck Computing and Data Facility}, Germany}\IEEEauthorrefmark{6}\\
Email: \IEEEauthorrefmark{1}gongq@ornl.gov}
}

\maketitle

\begin{abstract}
Data compression plays a key role in reducing storage and I/O costs. Traditional lossy methods primarily target data on rectilinear grids and cannot leverage the spatial coherence in unstructured mesh data, leading to suboptimal compression ratios. We present a multi-component, error-bounded compression framework designed to enhance the compression of floating-point unstructured mesh data, which is common in scientific applications. Our approach involves interpolating mesh data onto a rectilinear grid and then separately compressing the grid interpolation and the interpolation residuals. This method is general, independent of mesh types and typologies, and can be seamlessly integrated with existing lossy compressors for improved performance. 
We evaluated our framework across twelve variables from two synthetic datasets and two real-world simulation datasets. 
The results indicate that the multi-component framework consistently outperforms state-of-the-art lossy compressors on unstructured data, achieving, on average, a $2.3-3.5\times$ improvement in compression ratios, with error bounds ranging from $\num{1e-6}$ to $\num{1e-2}$. 
We further investigate the impact of hyperparameters, such as grid spacing and error allocation, to deliver optimal compression ratios in diverse datasets. 
\end{abstract}


\section{Introduction}
Unstructured meshes are widely utilized in finite element simulations due to their flexibility in representing complex geometries \cite{park2016unstructured,sakai2013parallel}. The resulting datasets consist of mesh topology and multi-variate data defined on mesh vertices. As advancements in HPC hardware enable higher resolution and faster simulation timesteps, the volume of generated data imposes immense demands on storage and transmission. For instance, a gas turbine jet engine simulation using GE's GENESIS code encompasses billions of mesh points \cite{ge-olcf}. A single variable in double-precision format can reach tens of terabytes, with post-analysis requiring time-series data across multiple time steps. This scenario necessitates the development of effective data reduction techniques that minimize information loss while preserving scientific integrity. 
Due to the high entropy in floating-point data produced by scientific computations, lossless compression techniques typically achieve limited compression ratios (e.g., less than $2\times$). \textit{Error-controlled lossy compression} has proven successful in scientific data compression by exploiting spatial and temporal redundancies and ensuring loss remains within user-prescribed bounds.    

In recent years, numerous lossy compressors for floating-point data have been developed, including MGARD \cite{gong2023mgard,ainsworth2018multilevel,ainsworth2019qoi}, SZ \cite{di2016fast,zhao2021optimizing,sz18,zhao2020significantly}, ZFP \cite{lindstrom2017error,lindstrom2014fixed}, SPERR \cite{li2023lossy}, and GAE \cite{lee2023nonlinear,lee2022error}. These compressors typically assume a regular grid layout and rely on block or stencil-based algorithms to convert data into small-magnitude coefficients amenable to compression. When applied to unstructured-mesh data, most of these techniques require serialization of node data followed by compression on 1D arrays. This serialization often leads to sub-optimal compression ratios due to overlooking data correlations within the mesh topology and the incoherent data distribution caused by the arbitrary connectivity through mesh cells. While value-based vertex traversal \cite{ren2023prediction} can mitigate some of these issues, it may slow down the compression and add overhead for storing the traversal graph. There are also methods that reduce unstructured data over their original mesh topology, but often transition data into entirely different formats \cite{el2021accurate}, which complicates post-analysis and error preservation, or are limited to specific mesh types or topology (e.g., tetrahedrons which can be gradually refined \cite{ainsworth2020multilevel}).

\begin{figure}
    \centering
    \includegraphics[width=0.48\textwidth]{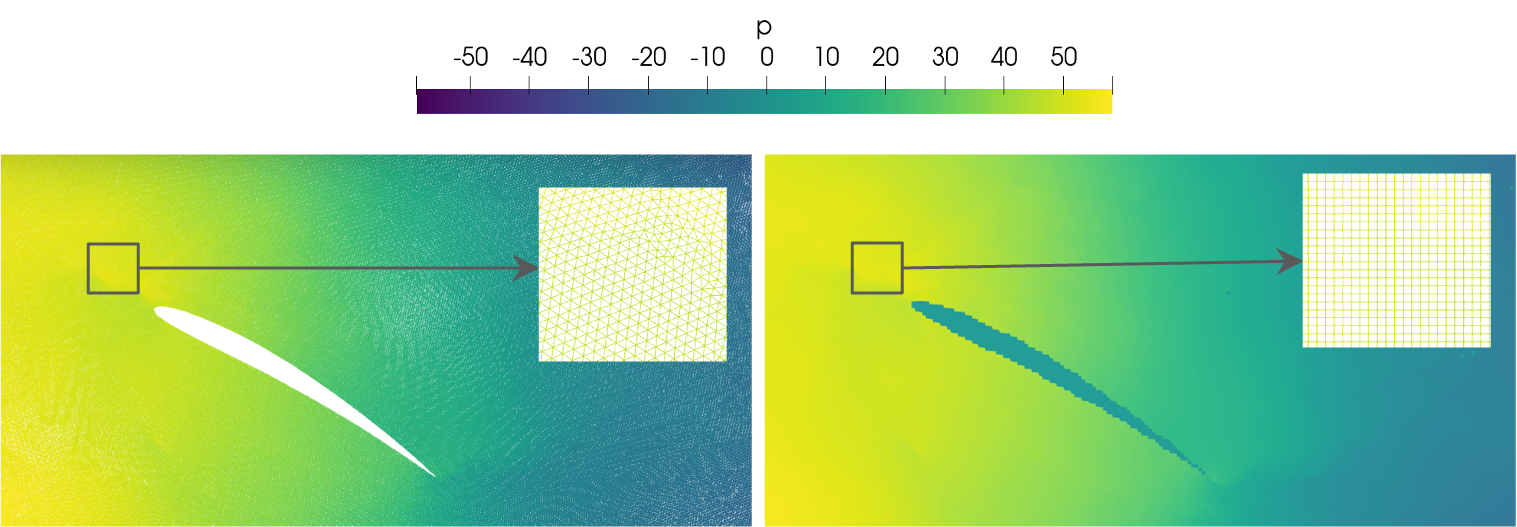}
    \caption{Example of a variable generated using OpenFoam on a 2D unstructured mesh (left) and its interpolation on a rectilinear grid (right). The airfoil blade is a hollow region in the left figure and interpolated as zeros in the right figure.} 
    \label{fig:motivation-interpolation}
\end{figure}

In this paper, we propose a generic approach that can be integrated with \textit{any} error-controlled lossy compressors, without requiring invasive code changes or custom mesh traversal, to significantly improve compression ratios for data defined on \textit{arbitrary} unstructured meshes. Our method approximates datasets defined over unstructured meshes onto rectilinear grids and performs \textit{multi-component} compression on the interpolated values on the rectilinear grid and the interpolation residuals on the original unstructured mesh vertices. 
To motivate our approach, we compare a 2D unstructured mesh and a 2D rectilinear grid. Figure~\ref{fig:motivation-interpolation} shows a variable (pressure) simulated by OpenFoam and its scatter-based interpolation on a 2D rectilinear grid. Using a number of nodes equivalent to $39\%$ of the mesh vertices, the resulting approximation faithfully conveys the properties of the original variable and produces low approximation errors when interpolated back to mesh vertices. These approximation errors, refereed as residuals in this paper, are more amenable to compression than the original data values, and the approximation on the rectilinear grid can also achieve high compression ratios by exploiting data correlations in high-dimensional space.  

Our framework can be summarized as follows. Given the vertices and associated data values of an unstructured grid and a user-prescribed error bound, our framework first constructs a rectilinear grid optimized for high compression ratios and low approximation errors. Next, we map the data values from the mesh vertices to the rectilinear grid nodes using a customized or user-supplied interpolation kernel and compute the residuals at the mesh vertices. Finally, we distribute the error bound and use error-controlled lossy compressors to independently reduce both the approximation and the residual components.

The salient aspects of our work are:
\begin{itemize}
    \item We develop a generic and effective framework to reduce unstructured mesh data based on mesh-rectilinear grid interpolation and multi-component compression. Existing error-controlled lossy compressors can be seamlessly integrated into our framework to enhance their performance in a rather non-disruptive manner. 
    \item Our approach does not require special types of meshes or typologies. Additionally, since the mesh-to-rectilinear-grid conversion is independent of the field values, the corresponding vertex mapping can be precomputed and reused across multiple variables and timesteps, which reduces storage overhead and accelerates computations.  
    \item The compression ratio and computational cost can be adjusted via the grid spacing and the error bounds used for compressing the grid approximation and residuals. The overhead in computing time varies with grid spacing, ranging from $26\%$ to $100\%$ under the parameter settings used in our experiments. Our approach achieves an average improvement in compression ratios of approximately $2.3-3.5\times$, and up to $14\times$, among three lossy compressors evaluated within our framework. 
\end{itemize}

The details of our approach follow a review of related work on error-controlled and unstructured data compression. We demonstrate the generality of the proposed framework through the integration with three state-of-the-art lossy compressors and evaluate performance across different variables taken from multiple synthetic and real applications simulated unstructured mesh datasets.  

\section{Related Work}
\label{sec:background}
Lossy compression reduces data by exploiting redundancies, achieving greater compression ratios than lossless compression at the cost of some accuracy. It's crucial that the data compressed using lossy methods meet numerical accuracy requirements to ensure the preservation of scientific integrity. State-of-the-art, error-controlled compressors fall into two categories: transformation-based approaches and prediction-based approaches. Transformation-based methods (e.g., MGARD \cite{gong2023mgard,liang2021mgard+,ainsworth2019qoi}, ZFP \cite{lindstrom2014fixed}, TTHRESH \cite{ballester2019tthresh}) utilize customized transformations to mitigate data redundancies. In contrast, prediction-based methods (e.g., ISABELA \cite{lakshminarasimhan2013isabela}, SZ \cite{liang2022sz3}) leverage diverse predictors to de-correlate data. Both types of methods produce coefficients of smaller magnitude that become zero after quantization, enabling compacted storage of the compressed data. 

Most existing lossy compressors are designed for rectilinear grid data. For example, ZFP divides data into non-overlapped blocks and transforms the floating-point data in each block into variable-length bit-streams independently. TTHRESH reduces dimensionality using techniques like Tucker and HOSVD, which rely on tensor decomposition on multidimensional grids. Similarly, all predictors implemented in the SZ family (e.g., Lorenzo, polynomial interpolation, and regression) require partitioning data into small blocks or sub-grids of different sampling rates. 

Few lossy compressors can directly handle unstructured mesh data. MGARD reduces data through orthogonal decomposition, requiring a hierarchical refinement to be performed among triangular/tetrahedral cells, which limits its use case \cite{ainsworth2020multilevel}. 
Ren et al. \cite{ren2023prediction} proposed a prediction-traversal approach that sequentially visits/compresses mesh nodes until an unpredictable node is encountered, then initiates a separate sequence from a new seed until all nodes are traversed. Due to the additional costs of storing traversal sequences, this approach delivers improvements only within limited error ranges and has a high computational overhead, with a throughput of less than 1MB/s. In contrast, most state-of-the-art lossy compressors (e.g., MGARD, SZ, ZFP) achieve throughputs of tens to hundreds of MB/s \cite{zfp-rd100,gong2023mgard}. 

Researchers also explored compressed sensing techniques for unstructured mesh data compression \cite{salloum2018optimal,salloum2015compressed}. However, like neural network-based compression models \cite{liu2019novel,zhang2021multi,lee2022error,lee2023nonlinear}, these methods suffer from unbounded errors and require tens to hundreds of iterations to reach a faithful reconstruction. 
Another set of related research involves the reduction of unstructured meshes. For example, El-Rushaidat et al. \cite{el2021accurate} aimed to convert unstructured data into rectilinear grids, while Berger and Rigoutsos \cite{berger1991algorithm} further explored cluster and adaptive mesh refinement (AMR) algorithms to reduce grids into non-uniform structure consist of fewer rectangular patches. Various studies have also addressed compressing data in AMR forms \cite{wang2023analyzing,luo2021zmesh,li2023lamp}. Our work is motivated by some mesh-to-grid data approximation techniques discussed in the aforementioned work but fundamentally differs in two key aspects: first, our primary objective is to reduce the storage cost rather than accelerate analytic tasks such as visualization; second, we aim to maintain the general structure of reduced data and mathematically preserve errors incurred during data compression. 

\section{Preliminaries}
\label{sec:error-proof}
\begin{figure}
    \centering
    \includegraphics[width=0.475\textwidth]{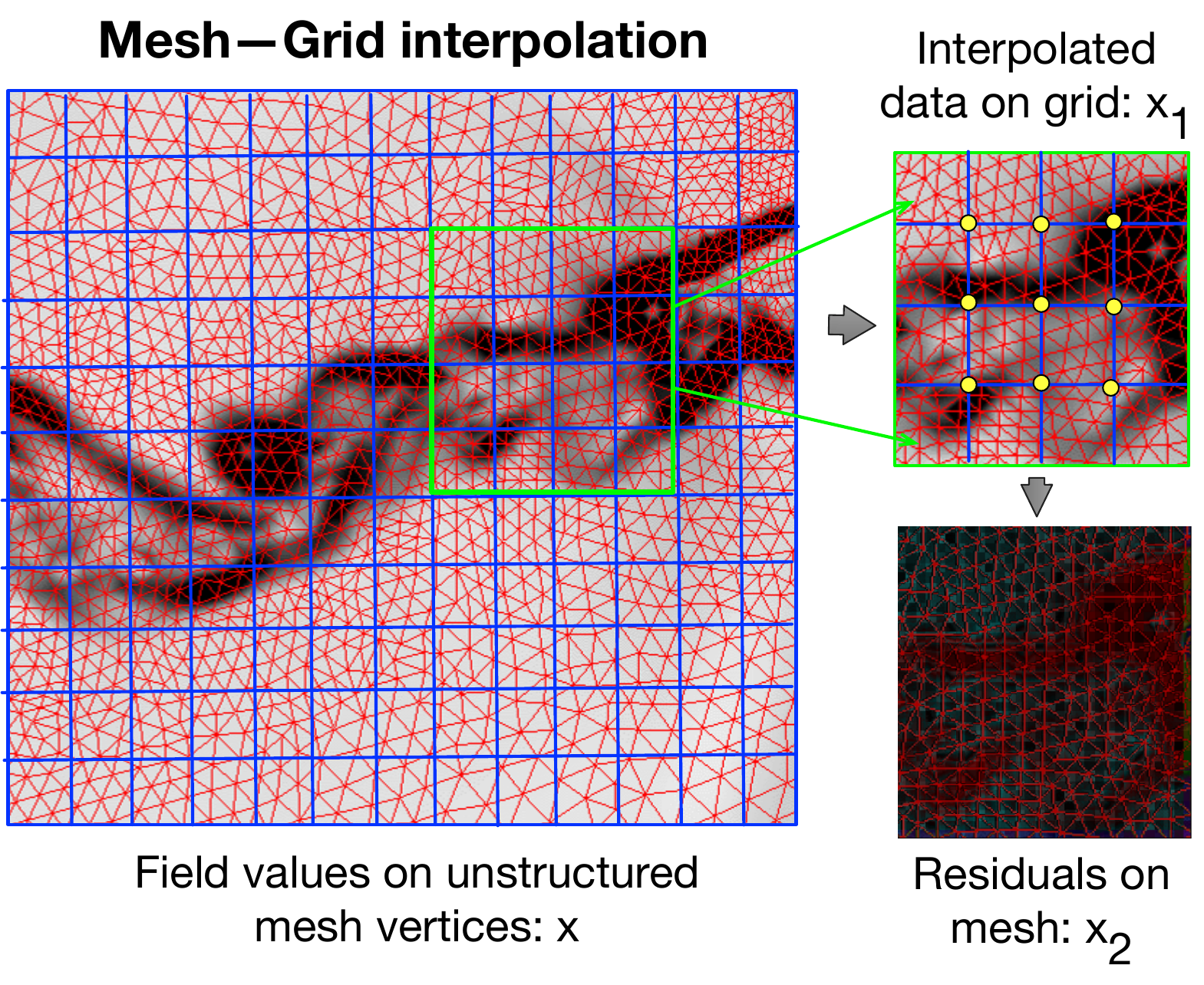}
    \caption{A multi-component compression for unstructured mesh data consists of building an interpolation on rectilinear grid and independent compression of grid interpolation and residuals on meshes. }
    \label{fig:multi-component}
\end{figure}

In this section, we provide an overview of the proposed framework and the proof of error control. As illustrated in Figure~\ref{fig:multi-component}, the key idea is to represent field values on mesh vertices using an approximation over a rectilinear grid and the approximation error (i.e., residuals). We denote the original field values on mesh vertices as $x$, the interpolated data on grid points as $x_1$, the residuals on mesh vertices as $x_2$, and the operator that interpolates $x_1$ back to mesh vertices as $g(.)$, where $g: \mathbb{R}^{n_1} \to \mathbb{R}^{n_2}$ is a multidimensional, multivariate function that maps the $n_1$ data points in the rectilinear grid to $n_2$ data points in the original grid. The original vertex values can then be reconstructed through $x = x_2 + g(x_1)$.    

Let $c(.)$ represent an error-controlled compressor that guarantees the error in the reconstructed data satisfies $\max |\epsilon| = \max |u-u'| \leq \tau$, where $\epsilon$ is the compression error, $\tau$ is the input error bound, and $u$ and $u'$ represent the original and lossy reduced data. Compression of the original data values on mesh vertices leads to an error $\max |\epsilon|=\max |x-x'| \leq \tau$, and compression for the grid approximation and residuals results in two other error components: $\max |\epsilon_1|=\max |x_1-x_1'| \leq \tau_1$, and $\max |\epsilon_2|=\max |x_2-x_2'| \leq \tau_2$. The reconstruction of $x$ using the lossy reduced components $x_1'$ and $x_2'$ can be captured by $x'= g(x_1')+x_2'$. 
For any linear interpolation function $g$, such that $g_i(x) = a^T x$, where $a$ is the interpolation coefficient, the error in the reconstructed data can be bounded by $(\sum_j |a_j| )\tau_1 + \tau_2$ according to basic linear algebra.
Hence, given a prescribed error bound $\tau$ on the unstructured mesh data $x$, 
we can compress the components of $x_1$ and $x_2$ under the error bound $\tau_1$ and $\tau_2$, respectively, and ensure the error in the reconstructed mesh data satisfies $|\epsilon| \leq \tau$ as long as $(\sum_j |a_j| )\tau_1 + \tau_2 \leq \tau$. 
In a special case where $g$ corresponds to nearest neighbor value or piece-wise linear interpolation of grid points, this requirement reduces to $\tau_1 + \tau_2 \leq \tau$ because $\sum_j |a_j| = 1$.


\section{Multi-component Unstructured Data Compression}
\subsection{Rationale of multi-component compression}
\label{sec:rationale}
In the previous section, we proved that compression for unstructured mesh data can be performed through independent compression of two components---the approximation on rectilinear grid and residuals---while ensuring the accuracy of recomposed data by adjusting the error bounds used for compressing individual components. We propose constructing the rectilinear grid based on the locations, rather than field values, of mesh vertices.     
The compression ratio (CR) of the proposed multi-component approach is captured as $\textrm{CR}=(c_1+c_2)/S$, where $c_1$ and $c_2$ are the compressed sizes of interpolation approximation and residual components, respectively, and $S$ is data original size. We exclude the grid and mesh layouts (e.g., node/vertices coordinates and cell connectivity) from the CR measurement as they remain static and can be reused across numerous timesteps and multiple variables sharing the same set of meshes.

Unlike vertices in unstructured meshes, whose indices often ignore the spatial coherence of field values, axis-aligned grids have implicit geometric coherence. Thus, lossy compression of approximation values on grids can better exploit the smoothness of scientific data in space and time. Additionally, assuming the interpolated grid values provide a faithful representation of the mesh data, the residual values will be small hence are more compressible than their original values. These points form the rationale for the proposed multi-component compression, suggesting that the combined size from compressing two components could be smaller than directly compressing unstructured mesh data. The main penalty of using multi-component compression lies in the overhead of compressing the grid interpolation and performing interpolation operations. Nevertheless, provided sufficiently larger compression ratios can be achieved, as demonstrated in Section~\ref{sec:experiment}, the reduction on I/O time and storage cost will likely compensate for the overhead in compression time. We also study the impact of grid sampling rate on approximate errors and the aggregated compression ratios in Section \ref{sec:hyperparameter}, providing guidance for users to make trade-offs.             

\subsection{Design overview}
\label{sec:overview}
\begin{figure*}
    \centering
    \includegraphics[width=0.95\textwidth]{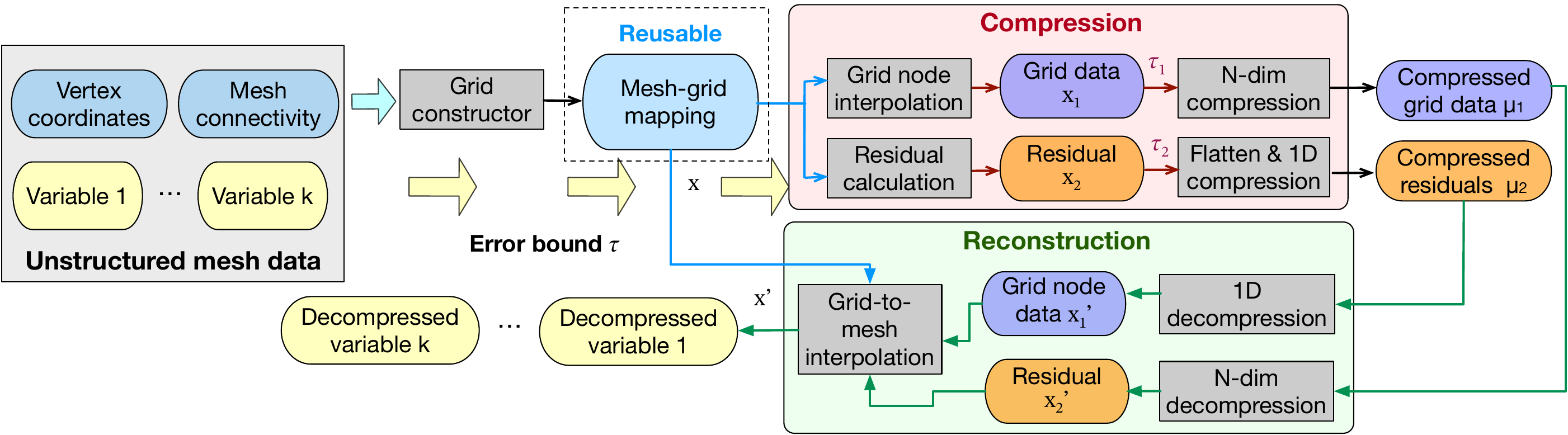}
    \caption{Workflow of the proposed multi-component, error-controlled compression and decompression algorithm for unstructured mesh data. We assume that there are multiple variables sharing the same set of meshes and the mesh remains static across different timesteps data, such that the mesh-grid mapping can be pre-computed to accelerate compression and decompression. }
    \label{fig:design-pipeline}
\end{figure*}

Figure \ref{fig:design-pipeline} illustrates the workflow of our multi-component compression and decompression for values on an unstructured mesh. 
The grid construction produces grid layouts and a mesh-grid mapping table, which can be reused across multivariate and timestep datasets to accelerate the interpolation operation $g(x_1)$ required for residual calculation and mesh data reconstruction. The mesh-to-grid interpolation can be performed through any customized algorithm, and we introduce a computationally light interpolation function in Section~\ref{sec:compression}. The residual values on mesh vertices are then calculated through $x-g(x_1)$, where $x$ is the original value at a mesh vertex. 

The multi-component framework splits the input error bound $\tau$ between $\tau_1$ and $\tau_2$, which are tolerance used for compressing the interpolated grid nodes and mesh residual values. The actual compression is conducted through error-controlled lossy compressors. The interpolated grid values are compressed in the N-dimensional Cartesian coordinate space, whereas the residuals are serialized into a 1D array for compression. The reconstruction follows an inverse procedure to compression, where the interpolation values (i.e., $x_1'$) and residuals (i.e., $x_2'$) are decompressed independently and recomposed to form the reconstructed values on mesh vertices. The entire pipeline is automated but allows users to optimize for individual cases by fine-tuning parameters such as the grid spacing and error distribution between grid interpolation and residuals. The hyperparameter tuning and its impact on compression performance will be explored in Section \ref{sec:hyperparameter}.        

\subsection{Grid Construction}
\label{sec:grid-selection}
The size of mesh cells is often nonuniform across computational space, with finer cells near boundaries and regions requiring a higher resolution. Our grid construction module consists of two steps: initializing grids with uniform spacing along each spatial dimension, followed by adaptive coarsening based on the spatial distribution of mesh vertices and the operator $g(x_1)$ used for grid-to-mesh interpolation. The module also outputs a vertex-wise index table $\{m_i\}$ recording the vertex-node mapping.    

The algorithm for initializing the uniform grid is presented in Algorithm \ref{alg:grid-init}. Generally, the grid construction module traverses meshes based on cell connectivity, measuring the distance between vertices and choosing the value at the bottom $k\%$ percentile (user-supplied parameter) along each spatial dimension as the grid spacing of the corresponding axis. Grid corners are assigned based on the minimum and maximum value of the mesh vertices along each dimension. As shown in the Algorithm \ref{alg:grid-init}, an upper bound on grid discretization $G_{\textrm{max}}$ may also be prescribed to avoid over discretization. 

\begin{algorithm}[h!]
\caption{\textsc{Grid\_Init}} \label{alg:grid-init} \footnotesize
\renewcommand{\algorithmiccomment}[1]{/*#1*/}
\begin{flushleft}
\textbf{Input}: mesh vertices coordinates \{$p_i$\}, connectivity \{$c_j$\}, dimension $D$, $G_\textrm{max}$ max number of nodes along each axis, $k\%$ile of vertex distance \\
\textbf{Output}: grid corners $\{gs_d\}$ and spacing \{min\_$p_d$, max\_$p_d$\}
\end{flushleft}
\begin{algorithmic} [1]
\STATE $\{\{s_i\}_d\}$, \{min\_$p_d$, max\_$p_d$\} $\gets$ \texttt{traverse\_mesh}(\{$p_i$\}, \{$c_j$\})  \COMMENT{obtain min/max, and vertex distance  along each dimension}
\FOR{$d = 1 \to D$}
    \STATE $\textrm{gs}_d$ $\gets$ \texttt{percentile}($k$, $\{s_i\}_d$) 
    \STATE $\textrm{gn}_d = (\textrm{max\_}p_d - \textrm{min\_}p_d) / \textrm{gs}_d$ 
    \WHILE{$\textrm{gn}_d \geq G_\textrm{max}$} 
        \STATE $k \gets k+\delta$
        \STATE $\textrm{gs}_d$ $\gets$ \texttt{percentile}($k$, $\{s_i\}_d$)
    \ENDWHILE
\ENDFOR
\RETURN $\{\textrm{gs}_d\}$, \{min\_$p_d$, max\_$p_d$\}
\end{algorithmic}
\end{algorithm}

We implemented a lightweight function $g(.)$ to minimize the computational overhead. For each mesh vertex $x[k]$, we approximate its value using the nearest grid node, $m_k$, such that the residual is derived as $x_2[k] = x[k] - x_1[m_k]$. Due to varied cell sizes, irregularly shaped boundaries, and hollow regions in the mesh geometry, some grid nodes may not be utilized during the residual and reconstruction computation, thus leading to resource waste. To address this, we adaptively coarsen the grid, as detailed in Algorithm \ref{alg:grid-coarsen}. The grid coarsen module first traverses the mesh vertices to find their closest grid nodes. This traversal has two objectives: (1) building a grid-to-mesh mapping which can be reused to accelerate residual calculation and data recomposition, and (2) flagging grid nodes that are unvisited in back interpolation. Next, the algorithm scans along each grid axis, checking for unvisited arrays/planes.             

\begin{algorithm}[h!]
\caption{\textsc{Grid\_Coarsen}} \label{alg:grid-coarsen} \footnotesize
\renewcommand{\algorithmiccomment}[1]{/*#1*/}
\begin{flushleft}
\textbf{Input}: mesh vertex coordinates \{$p_i$\}, grid layout $S$, dimension $D$ \\
\textbf{Output}: mesh-grid node mapping $\{m_i\}$, updated $S$
\end{flushleft}
\begin{algorithmic} [1]
\STATE $G$\_node.\textit{visited} $\gets$ \texttt{False}
\STATE \COMMENT{loop through $n_v$ mesh vertices}
\FOR{$i = 1 \to n_v$} 
    \STATE $m_i$ $\gets$ \texttt{MAP}($p_i$, $S$) \COMMENT{mapping varies with $g(.)$}
    \STATE $G$\_node[$m_i$].\textit{visit} $\gets$ \texttt{True}
\ENDFOR
\STATE \COMMENT{loop through grid nodes}
\FOR{$d = 1 \to D$}
    \FOR{$k$ in coord[$d$]}
        \STATE planes.\textit{visited} $\gets$ \texttt{check\_visit}($G$\_node.\textit{visited}, $S$, $k$) \COMMENT{check if any node in the plane forms by the rest dimensions was visited}
        \IF{planes.\textit{visited} == \texttt{False}}
        \STATE update $S$, $\{m_i\}$ 
        \ENDIF
    \ENDFOR
\ENDFOR
\RETURN $\{m_i\}$, $S$
\end{algorithmic}
\end{algorithm}

Since the AMR-type data cannot be handled by most lossy compressors, coarsening must be performed on all nodes along the corresponding dimensions within the coarsened range. For irregular-shaped meshes, they may yield grids with ``blank'' regions/nodes that cannot be eliminated due to the continuity along grid axes. Hence, our framework implements a variation algorithm where the component $x_1$ contains only grid nodes visited in Algorithm \ref{alg:grid-coarsen}. These nodes are treated as \textit{seeds} for vertices in surrounding regions to interpolate residuals and then serialized for compression. 
Our framework automatically switches to the variation implementation when the percentage of $G$\_node.\textit{visit} fails to reach a given threshold.      

\subsection{Compression}
\label{sec:compression}
Algorithm \ref{alg:compression} details the multi-component compression module, which consists of three parts: grid values interpolation, residual calculation, and independent compression of interpolated values and residuals. We denote the interpolation function used for approximating grid values as $f(.)$. Note that the objective of our research is to maximize the compression ratio of unstructured data under a numerical error bound, rather than to produce the best ``visualization''. The function $f(.)$ that produces the smoothest approximation may not lead to the smallest magnitude for mesh residuals since the back interpolation $g(.)$ used for residual calculation must be a linear function to ensure error preservation. Considering the overhead incurred to compression throughput, our framework implements two operators for $f(.)$, one based on linear interpolation and another based on cluster. The cluster-based function takes the mean of $n$ vertices data closest to node $j$ (i.e., $x\in C_j$) as the interpolated value, i.e., $x_{1}[j] = {\sum_{x\in C_j}^n x[i]}/{n}$. Despite its simplicity, $x_1[j]$ minimizes the residuals as $\sum_{x\in C_j}(x[i]-x_{1}[j])^2$ is smallest when $x_{1}[j]$ is the mean value of the $x[i]$ assigned to the cluster $C_j$. Users can replace $f(.)$ using other customized interpolation functions. 
The residual computation and independent compression---$\mathbb{C}(x_1)$ and $\mathbb{C}(x_2)$---are described in Section \ref{sec:overview}. Here, $\mathbb{C}$ can be any error-controlled lossy compressor.
\begin{algorithm}[h!]
\caption{\textsc{Multi-component compression}} \label{alg:compression} \footnotesize
\renewcommand{\algorithmiccomment}[1]{/*#1*/}
\begin{flushleft}
\textbf{Input}: mesh-grid mapping \{$m_i$\}, mesh vertices value $\{x_i\}$, grid layout $S$ \\
\textbf{Output}: compressed residual values $\mu_2$, compressed grid values $\mu_1$
\end{flushleft}
\begin{algorithmic} [1]
\STATE $x_1 \gets 0$
\STATE \COMMENT{Obtain grid interpolation values}
\FOR{$i = 1 \to n_v$} 
    \STATE $x_1[m_i] \gets f(x_i, x_1[m_i])$  
\ENDFOR
\STATE $x_2 \gets x - g(x_1)$
\STATE \COMMENT{multi-component compression}
\STATE $\mu_1 = \mathbb{C}(x_1)$ \COMMENT{$\mathbb{C}$ as the compression function}
\STATE $\mu_2 = \mathbb{C}(x_2)$
\RETURN $\mu_1, \mu_2$
\end{algorithmic}
\end{algorithm}


\subsection{Reconstruction}
Reconstruction follows the inverse procedure of compression, as detailed in Algorithm \ref{alg:reconstruction}. Similar to multi-component compression, grid approximation and mesh residual values are decompressed independently. Data values are recomposed back to the original mesh vertices through: $x' = g(x_1') + x_2'$. Errors in the recomposed mesh data are guaranteed to stay within the user-prescribed input tolerance, as we proved in Section \ref{sec:error-proof}.  

\begin{algorithm}[h!]
\caption{\textsc{Multi-component reconstruction}} \label{alg:reconstruction} \footnotesize
\renewcommand{\algorithmiccomment}[1]{/*#1*/}
\begin{flushleft}
\textbf{Input}: compressed residual values $\mu_2$, compressed grid values $\mu_1$, grid layout $S$, mesh-grid mapping \{$m_i$\} \\
\textbf{Output}: reconstructed mesh values $x'$
\end{flushleft}
\begin{algorithmic} [1]
\STATE $x_1' \gets \mathbb{D}(\mu_1)$ \COMMENT{$\mathbb{D}$ as the decompression function}
\STATE $x_2' \gets \mathbb{D}(\mu_2)$
\FOR{$i = 1 \to n_v$} 
    \STATE $x'_i \gets g(x'_1[m_i]) + x'_{2i}$
\ENDFOR
\RETURN $x'$
\end{algorithmic}
\end{algorithm}

\section{Experiments}
\label{sec:experiment}
\subsection{Datasets and evaluation metrics}
\begin{figure*}[t]
  \centering
     \includegraphics[width=\linewidth]{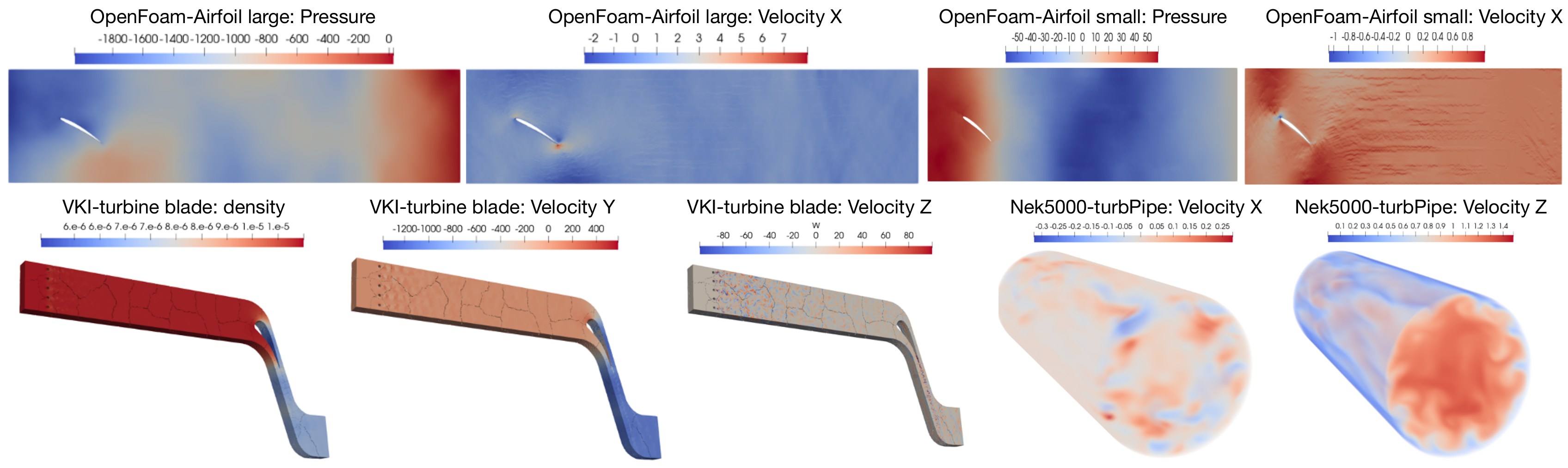}
  \caption{Visualization of the benchmark mesh data}    
    \label{fig:vis_mesh}   
\end{figure*}
In this section, we evaluate our multi-component compression framework using 12 variables captured from two synthetic datasets and two real-world simulation datasets. Detailed specifications are presented in Table~\ref{tab:dataset}, with selective visualization shown in Figure \ref{fig:vis_mesh}. The synthetic datasets were generated by OpenFoam \cite{jasak2009openfoam} through the PISO algorithm \cite{jang1986comparison}, which solves the pressure-velocity calculation for the Navier-Stokes equations. We simulated turbulent flow for a 2D NACA airfoil validation case on a small and large sets of meshes and initialization parameters, denoted as airfoil-small and airfoil-large. 
The Nek5000-turbPipe dataset was obtained from a simulation of fully-developed turbulent pipe flow \cite{rezaeiravesh2019statistics} generated by Nek5000 \cite{merzari2020toward}, a computational fluid dynamics code. The VKI dataset was obtained from a simulation of the VKI (von Karman Institute) gas turbine blade cascade test case \cite{arts1992aero} generated by GE's GENSIS software. 

We utilize three state-of-the-art error-controlled lossy compressors (detailed in Table \ref{tab:compressor}) within our framework to demonstrate the enhancement and generality. As reviewed in Section \ref{sec:background}, these compressors are designed for compressing scientific data on rectilinear grids, requiring mesh data to be serialize, usually based on their original order in the files, before compression \cite{liang2022toward,liang2020toward}. We denote this 1D serialization-based compression as \textit{default}. 

Throughout the evaluation, we demonstrate the impact of different hyper-parameters and how compression ratios can be significantly improved under our multi-component framework. We define the following metrics: (1) Compression ratio (CR): defined in Section \ref{sec:rationale}. (2) Achieved compression error $\epsilon$: errors measured in reconstructed data, required to be less than the input compression error bound $\tau$. We choose the relative $l2$ error, denoted as $\epsilon = \sum{\sqrt{(x-x')^2}} / \sum{\sqrt{x^2}}$, as the error metric, where $\sum{\sqrt{x^2}}$ represents the norm of the original data. (3) Improvement ratio: the ratio of the compression ratios achieved by our multi-component framework to the default approach implemented with the same lossy compressor. (4) Throughput overhead: the \textit{extra} time spent using our multi-component framework comparing to the default method.         

The experiments were conducted on the Frontier supercomputer \cite{atchley2023frontier}. Each Frontier compute node consists of a 64-core AMD 3rd Gen EPYC CPU with access to 512 GB of DDR4 memory. All source code was writeen in C/C++. Although most lossy compressors tested in this work support multi-threading or GPU acceleration, we tested only the single core CPU execution mode as code parallelism is irrelevant to our evaluation metrics. 

\newcolumntype{C}[1]{>{\centering\arraybackslash}p{#1}}
\begin{table}[h!]
  \centering
  \begin{tabular}{C{1.5cm}|C{0.6cm}|C{2cm}|C{1.5cm}}
    \hline
    dataset 1 & dim & attributes & \# of vertices \\
    \hline
    Nek5000-turbPipe & 3D & velocities & 1,451,520 \\
    \hline
    VIK turbine blade & 3D & pressure, density, Velocities & 209,121,200 \\
    \hline
    OpenFoam airfoil-small & 2D & pressure, velocities & 527,904 \\
    \hline
    OpenFoam airfoil-large & 2D & pressure, velocities & 1,994,776 \\
    \hline
  \end{tabular}
  \caption{Data sets used for benchmark}
  \label{tab:dataset}
\end{table}

\begin{table}[h!]
  \centering
  \begin{tabular}{C{1.4cm}|C{1cm}|C{4.75cm}}
    \hline
    Compressor & Version & GitHub repository \\
    \hline
    MGARD & 1.5.2 & https://github.com/CODARcode/MGARD\\
    \hline
    SZ3 & 3.1.8 & https://github.com/szcompressor/SZ3 \\
    \hline
    ZFP & 1.0.1 & https://github.com/LLNL/zfp \\
    \hline
  \end{tabular}
  \caption{Compressors used in our multi-component compression framework}
  \label{tab:compressor}
\end{table}

\subsection{Hyperparameter optimization}
\label{sec:hyperparameter}

\begin{figure*}[t]
  \centering
  \begin{subfigure}[b]{0.49\linewidth}
    \centering
     \includegraphics[width=\linewidth]{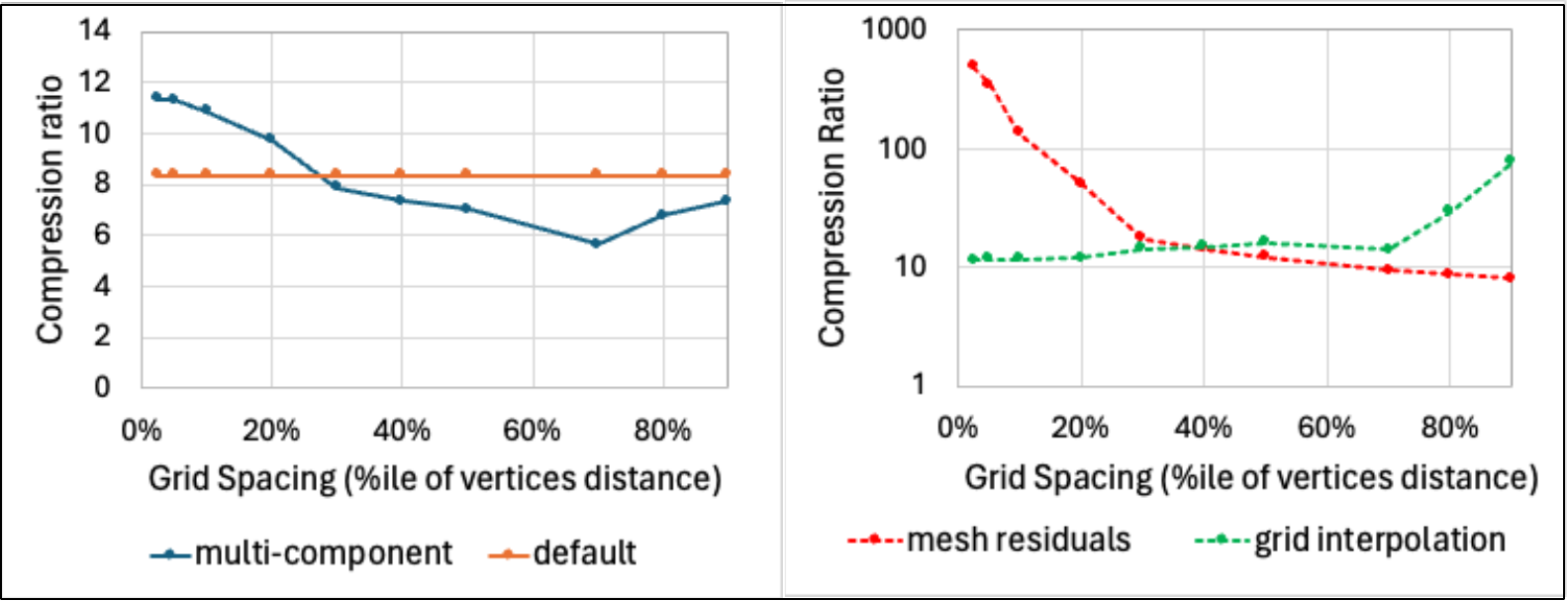}
     \caption{Impact of grid spacing using Nek5000 velocity data}
     \label{fig:impact_GridSel_nek}   
  \end{subfigure}
\begin{subfigure}[b]{0.49\linewidth}
  \centering
  \includegraphics[width=\linewidth]{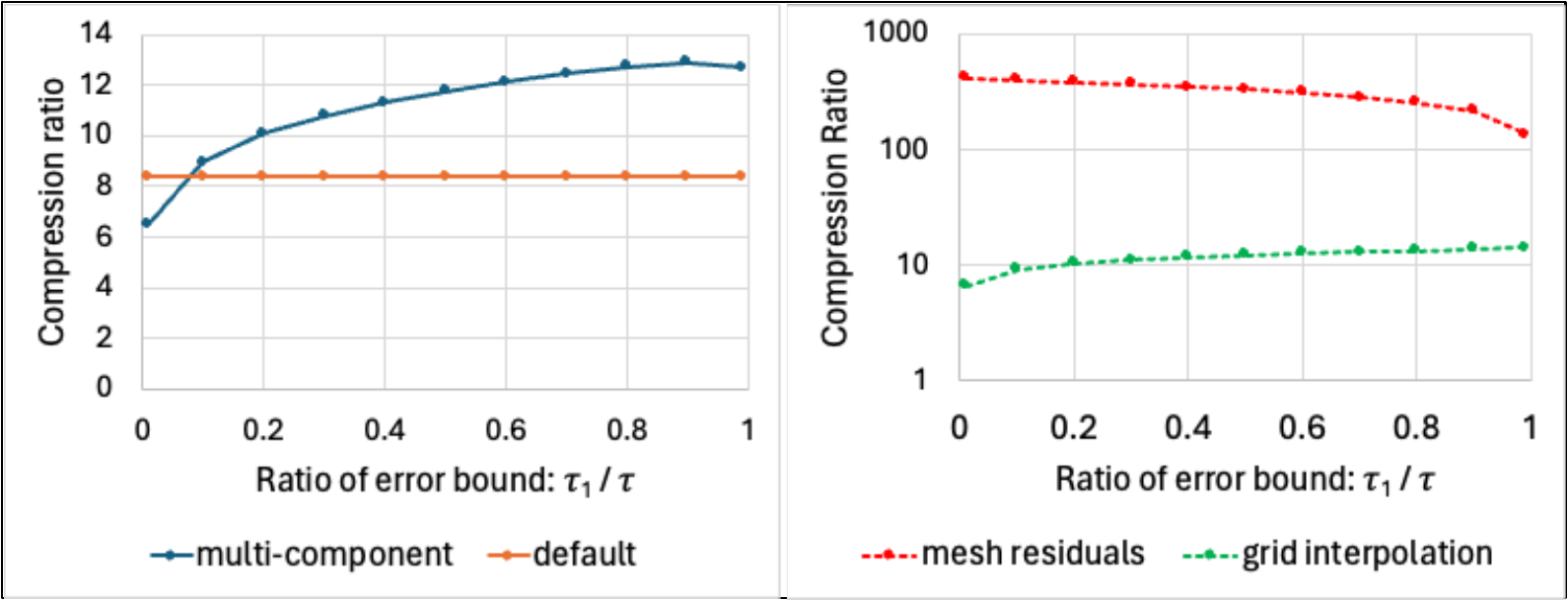}
  \caption{Impact of error allocation using Nek5000 velocity data}
  \label{fig:impact_ebRatio_nek}
  \end{subfigure}  
  \hfill
\begin{subfigure}[b]{0.49\linewidth}
  \centering
  \includegraphics[width=\linewidth]{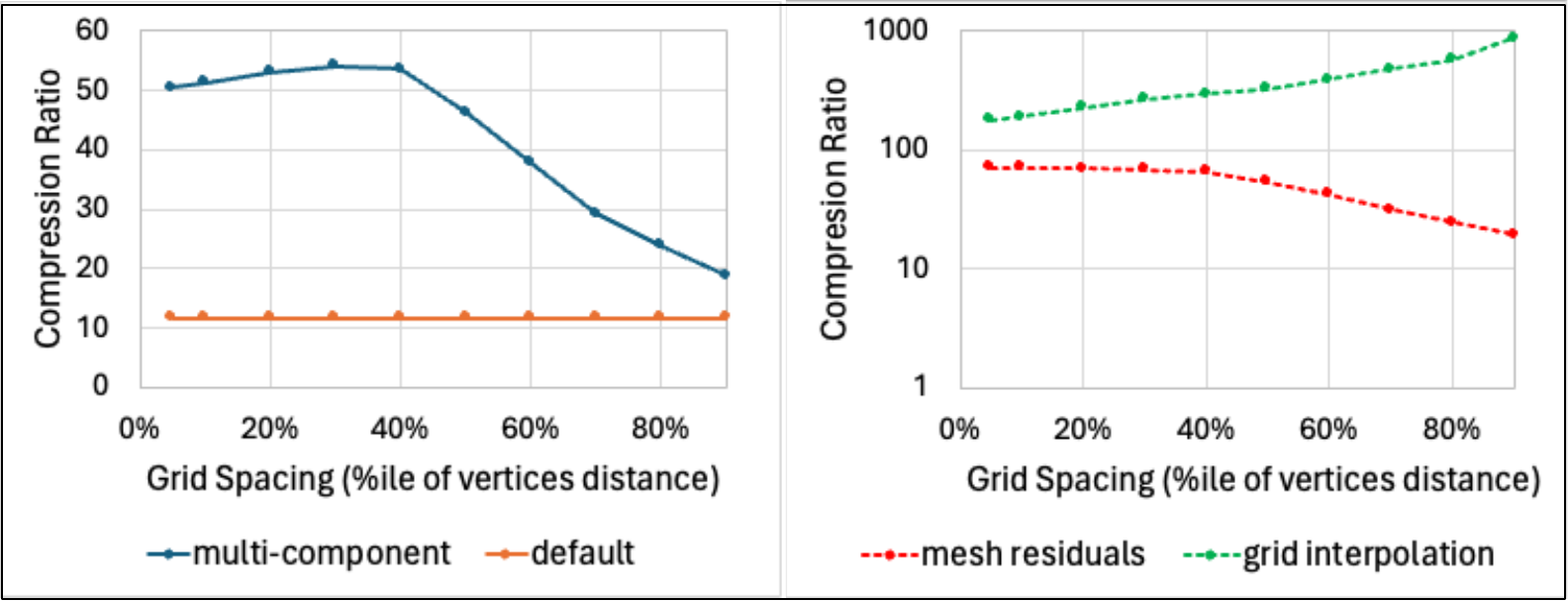}
  \caption{Impact of grid spacing using VKI pressure data}
  \label{fig:impact_GridSel_vki}
  \end{subfigure}  
\begin{subfigure}[b]{0.49\linewidth}
  \centering
  \includegraphics[width=\linewidth]{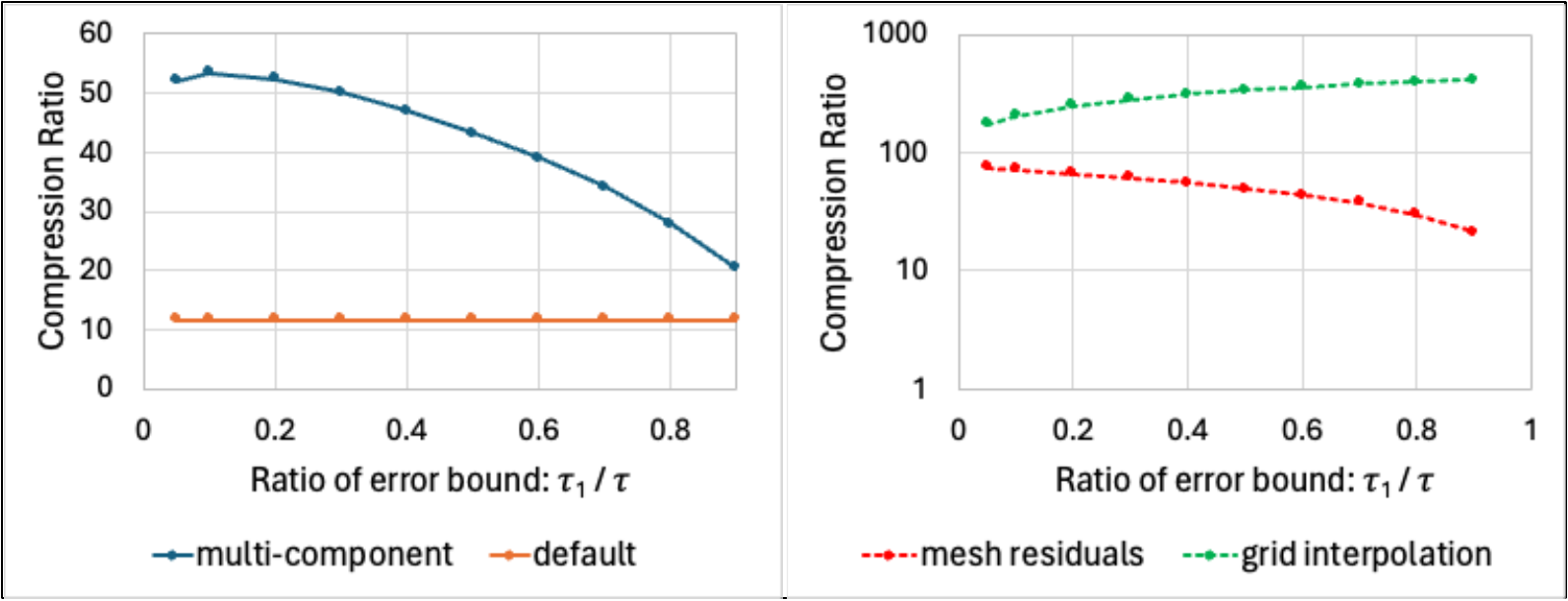}
  \caption{Impact of error allocation using VKI pressure data}
  \label{fig:impact_ebRatio_vki}
  \end{subfigure}  
  \caption{Impact of hyper-parameters on achieved compression ratios: demonstrated using data exhibiting different scales of smoothness.}   
  \label{fig:hyperparameter}
\end{figure*}
We begin by exploring the impact of grid spacing on multi-component compression ratios. The grid spacing is the node distance along each spatial dimension, which is selected based on the percentile ($\%$ile) of the vertex distance, as discussed in Algorithm \ref{alg:grid-init}. It represents the \textit{finest} scale in grid approximation. We evaluate parameter settings using the \textit{highly fluctuated} velocity data from a fully-developed turbulent pipe flow simulated by Nek5000 and the pressure data from a \textit{relatively smooth} region in a VKI turbine blade simulation.  
Figure~\ref{fig:hyperparameter} illustrates the compression ratios under different parameter settings, with combined compression ratios plotted on the left and the individual compression ratios for grid interpolation and residual data on the right in each subfigure. 
As shown in Figure~\ref{fig:impact_GridSel_nek} and \ref{fig:impact_GridSel_vki}, as the grid spacing coarsens, the compression ratio of grid interpolation data rises due to the reduced number of grid nodes. Meanwhile, the compression ratio of interpolation residuals decreases due to increased approximation errors. The combined compression ratios peak at grid spacing equal to around $30-40\%$ of the vertex distance percentile for VKI pressure data and at $1\%$ of the vertex distance percentile for Nek5000 data. The turbulent Nek5000 data requires fine-grained grid representation, or it would yield significant interpolation errors (i.e., mesh residuals). The VKI pressure data, in comparison, are relatively smooth, 
thus can afford to interpolate on coarser grids.

Next, we fix the grid spacing at the $1\%$ percentile and $20\%$ percentile for the Nek5000 and VIK datasets, respectively, and study the impact of error allocation in Figure \ref{fig:impact_ebRatio_nek} and \ref{fig:impact_ebRatio_vki}. Our implementation requires $\tau_1 + \tau_2 \leq \tau$, where $\tau$ is the requested error bound on mesh data, $\tau_1$ and $\tau_2$ are error bounds used for compressing the grid interpolation and mesh residuals, respectively. As the $\tau_1$ increases, the compression ratio of grid data becomes larger, and the compression ratio of interpolation residual drops. However, the combined compression ratios for Nek5000 and VKI datasets exhibit opposite trends because the combined ratio is upper bounded by the compressibility of grid interpolation for Nek5000 data and by interpolation residuals for VKI data. 

\subsection{Benchmark Datasets Evaluation}
\label{sec:benchmark_cr}
\begin{figure}[t]
  \centering
     \includegraphics[width=\linewidth]{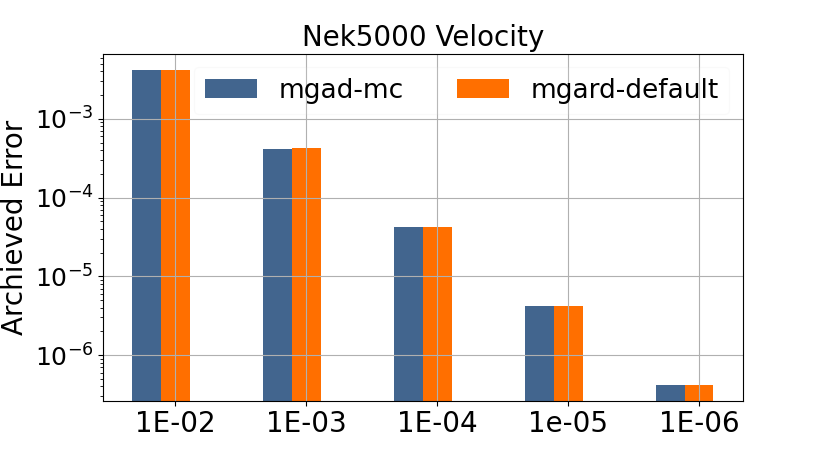}
  \caption{Compression error analysis, with x-axis representing the requested/input error bound and y-axis representing the archived relative root-of-mean-square error measured in lossy compressed data. The evaluation was conducted using Nek5000-turbPipe's velocity field.}     
  \label{fig:err_eval}   
\end{figure}

To verify that our framework meets error bounds, we use the Nek5000-turbPipe velocity field and plot in Figure \ref{fig:err_eval} the measured error $\epsilon$ as a function of input error bound $\tau$. We denote the data compressed through serialization by ``-default'' and the proposed multi-component approach by ``-mc'', with both approaches using MGARD as the underlying compressor. The evaluation shows that the errors in multi-component compressed data stay below the prescribed error bounds and are comparable to the ones derived through standard serialization-based approach.

\begin{figure*}[t]
  \centering
    \includegraphics[width=\linewidth]{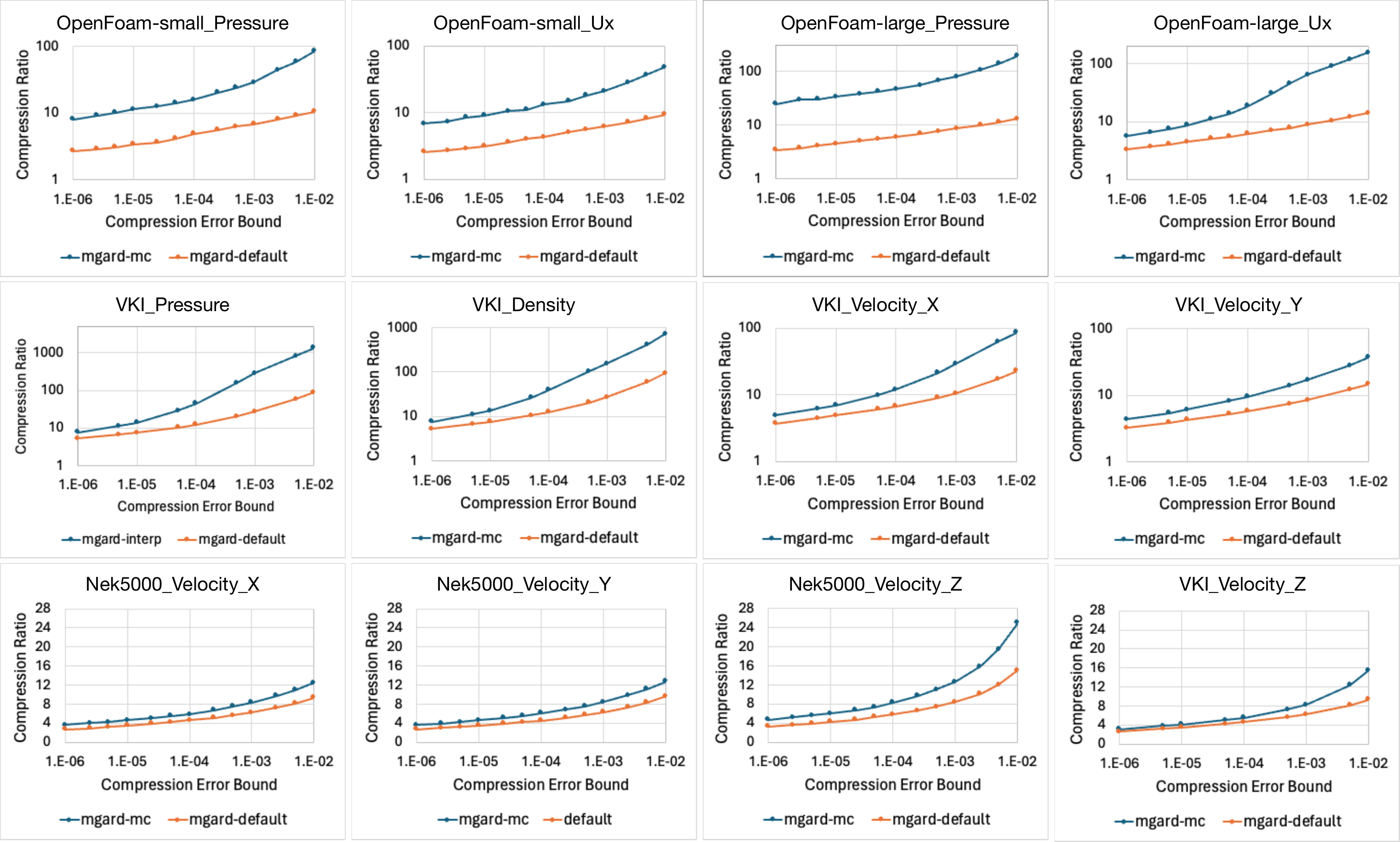}
  \caption{Compression ratios comparison between default and the proposed multi-component approach, exemplified using MGARD as the underlying lossy compressor.}   
  \label{fig:improvement_mgard}
\end{figure*}

Using MGARD as the underlying compressor, we plotted in Figure \ref{fig:improvement_mgard} the compression ratios as a function of $\tau$ for 12 variables in four unstructured datasets. We chose a grid spacing of $10\%$, $40\%$, $30\%$ percentile, and $\tau_1/\tau=0.99$, $0.15$, $0.9$ for all variables in Nek5000, VKI, and two OpenFoam airfoil datasets, respectively. Different variables from the same the dataset (e.g., density, pressure, and three velocities in VKI turbine blade) share one set of parameter settings. In this way, the cost of grid construction can be amortized, not counting towards overhead. Several general trends were noted. First, the multi-component compression (blue curves) almost always outperforms the serialization approach (orange curves). Second, the improvement ratio is not constant but generally larger at high error ranges. This is expected as large error bounds would quantize more residual data into zeros, leading to higher improvements in combined compression ratios. Third, the improvement ratio is typically higher for more ``compressible'' data, where a compression ratio of $10\times$ or more can be achieved using the serialization approach under a relative small error bound. This usually indicates strong spatial correlation, which our multi-component compression can better leverage. For example, the velocity Y vs. velocity Z values in VKI turbine blade dataset and the velocity Z vs. velocity X values in Nek5000 turbPipe dataset.

\begin{figure*}[t]
     \centering
     \includegraphics[width=\linewidth]{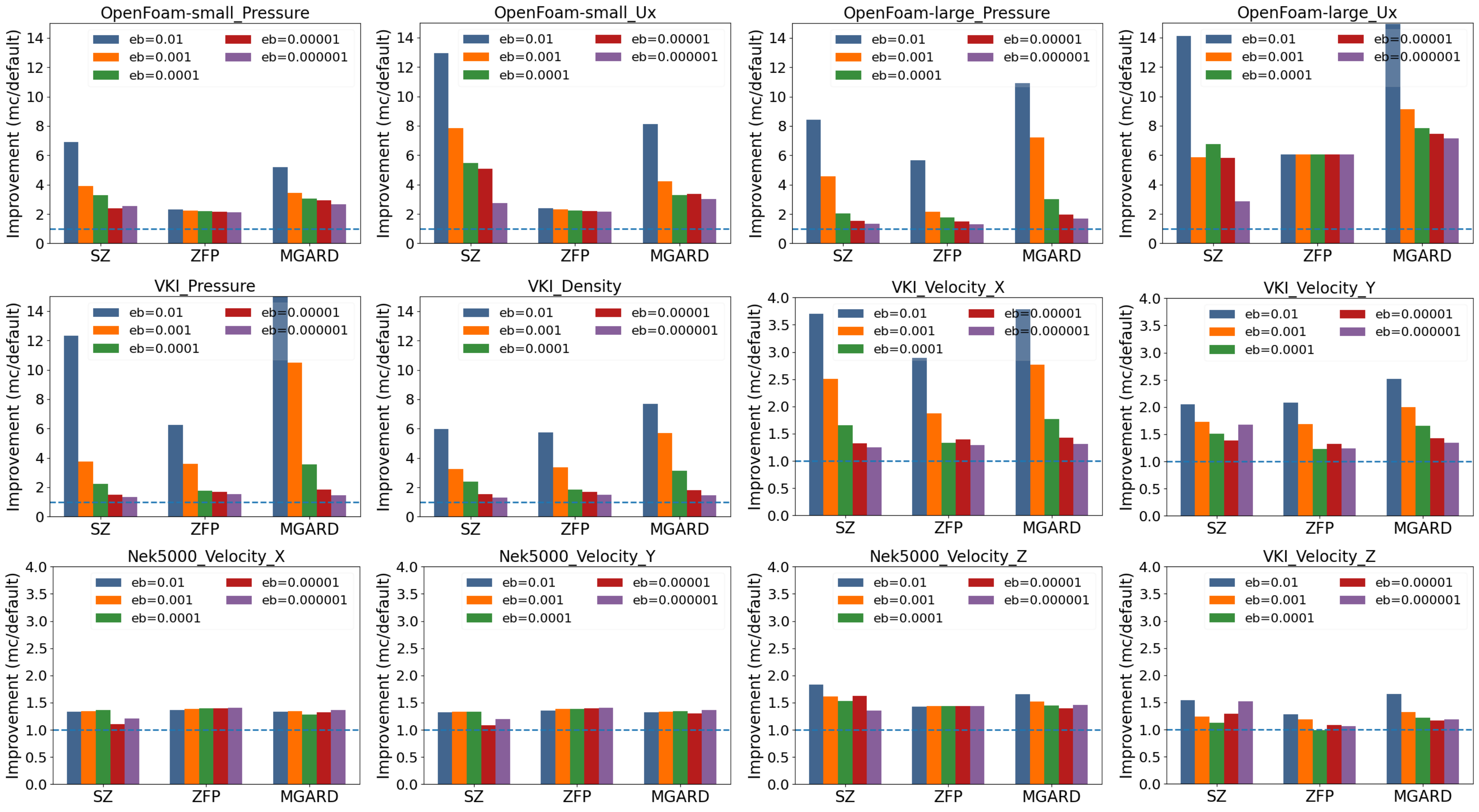}
  \caption{Improvement ratios achieved by the proposed multi-component approach, demonstrating using three lossy compressors across different error bounds, compared to the serialization-based compression conducted through the same underlying compressors.}     
  \label{fig:improvement_overall}   
\end{figure*}

Finally, we plotted in Figure \ref{fig:improvement_overall} the improvement ratios achieved for three lossy compressors---SZ, MGARD, and ZFP---using our multi-component framework. The improvement ratio for three compressors exhibits a trend similar -- the multi-component implementation delivers larger improvement ratios at high error ranges, but their level of improvement varies across datasets. 
We suspect the distinction is due to different de-correlation and quantization functions used by the three underlying compressors. 
Across the entire set of variables and error bounds, our multi-component compression consistently outperforms the serialization-based implementation, with improvement ratios $\geq 1$ for all three lossy compressors, except for ZFP at $\tau=\num{1e-4}$ with the velocity Z data in VKI turbine blade dataset. We notice that VKI-velocityZ data field is extremely difficult to compress, with values following almost a random distribution with a mean of zero, as illustrated in Figure \ref{fig:vis_mesh}. ZFP exploits data redundancies through breaking data into $4^d$ blocks and performing orthogonal transformations. Given that the compression of residual data is conducted in 1D rather than the $4\times4\times4$ space preferred by the transformation algorithm, ZFP is expected to obtain less significant improvement than the other two compressors. Across the 12 mesh variables and five error tolerance ranges, MGARD, SZ, and ZFP achieve an average improvement ratio of $3.5\times$, $3.1\times$, and $2.3\times$ respectively.    

\subsection{Runtime overhead}
\label{sec:runtime}
\begin{figure}[t]
  \centering
     \includegraphics[width=0.9\linewidth]{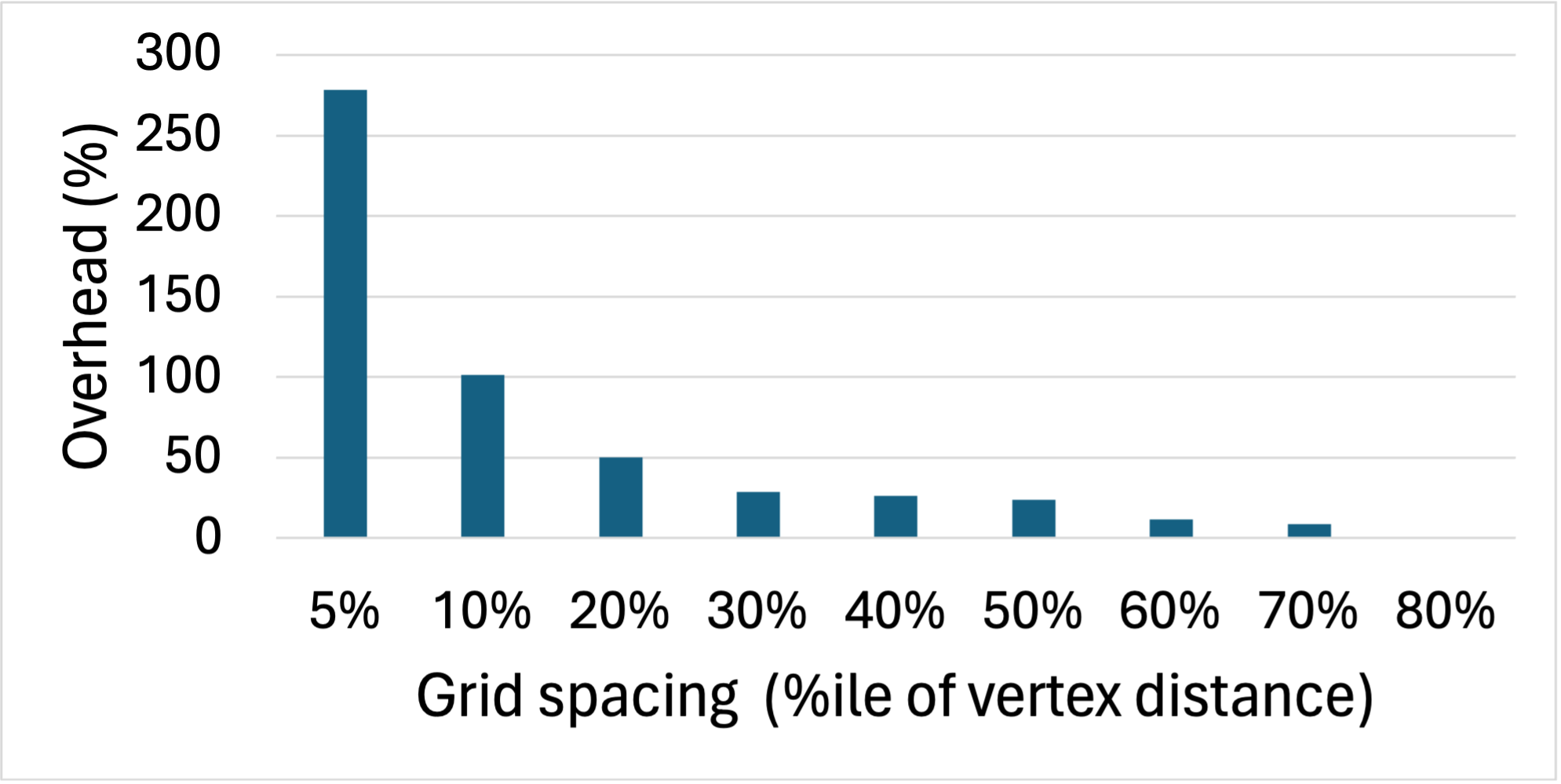}
  \caption{Overhead of throughput performance as a function of the grid spacing.}     
  \label{fig:throughput}   
\end{figure}

Our evaluation excludes the cost of mesh-to-grid translation, as the resulting map can be reused across multiple timesteps and data fields, providing that the mesh structure remains unchanged. Hence, the overhead of the multi-component compression comes from the compression of grid approximation, mesh-to-grid data interpolation (i.e., $f(.)$), and the back interpolation (i.e., $g(.)$) for residual calculation. Given $f(.)$ and $g(.)$, the computation overhead is closely linked to the number of grid points used for interpolation, which is controlled by the hyperparameter of grid spacing. We therefore evaluate the overhead of multi-component compression as a function of grid spacing and plot the results in Figure \ref{fig:throughput}. We denote the \textit{overhead} as $(\textrm{T}_\textrm{mc} - \textrm{T}_\textrm{default}) / \textrm{T}_\textrm{default}$, where $\textrm{T}_\textrm{default}$ represents the time spent on compressing using the same underlying compressor through the serialization approach. For grid spacing of $10\%, 30\%$, and $40\%$, which are parameters used for the evaluations conducted in Section \ref{sec:benchmark_cr}, the computational overhead ranged from $26\%$ to $101\%$. This overhead can be compensated considering the speedup for I/O and storage savings with the greatly reduced data size obtained using our approach. Users can also trade off the compression ratios for smaller computational overhead when throughput is a major concern.  

\section{Conclusion}
\label{sec:conclusion}
In this paper, we develop a multi-component, error-bounded compression framework tailored for scientific data on unstructured meshes, addressing the limitations of existing methods that primarily target rectilinear grid data. Our framework interpolates mesh data onto a rectilinear grid, then independently compresses both the grid representation and the approximation errors on meshes (i.e., residuals). Reconstruction is achieved by combining the decompressed grid data and mesh residuals at the original mesh vertices. Our solution is effective and general, seamlessly integrating into existing I/O and compression pipelines to enhance the compression of unstructured mesh data with minimal code changes. Extensive experiments using OpenFoam synthesized and real simulation datasets demonstrate that our approach achieves, on average, a $2.3-3.5\times$ improvement in compression ratios when used with three state-of-the-art lossy compressors, within an error bound range of $\num{1e-6}-\num{1e-2}$. In the future, we plan to investigate potential improvements offered by using different interpolation functions (i.e., $f(.)$ and $g(.)$ as discussed in the paper) and assess their impact on throughout. We will also evaluate the combined time spent on compression and the I/O of reduced data.  

\section*{Acknowledgment}
This research was supported by the SIRIUS-2 ASCR research project, the Scientific Discovery through Advanced Computing (SciDAC) program, specifically the RAPIDS-2 SciDAC institute, and the GE-ORNL CRADA data reductoin project. 
This research used resources of the Oak Ridge Leadership Computing Facility, which is a DOE Office of Science User Facility.

\bibliographystyle{IEEEtran}
\bibliography{reference}

\end{document}